\documentstyle[preprint,aps,epsfig]{revtex}
\begin{document}

\tightenlines
%\twocolumn[\hsize\textwidth\columnwidth\hsize\csname @twocolumnfalse\endcsname
%\bibliographystyle{plain}

%\vskip2pc] 
%\narrowtext

\draft

\title{Heat kernel of integrable billiards in a magnetic field.}
\author{R.~Narevich, D.~Spehner\cite{spehner}   
and E.~Akkermans\\ Department of Physics, 
Technion, 32000 Haifa, Israel}

\maketitle

\begin{abstract}
We present analytical methods to calculate the magnetic response of 
non-interacting electrons constrained to a domain with boundaries and
submitted to a uniform magnetic field.
Two different methods of calculation are considered - one involving the large
energy asymptotic expansion of the resolvent (Stewartson-Waechter method)
is applicable to the case of separable systems, and  
another based on the small time asymptotic behaviour of the heat kernel
(Balian-Bloch method). 
Both methods are in agreement with each other but differ from the result 
obtained previously by Robnik. Finally, the Balian-Bloch multiple scattering 
expansion is studied and the extension of our results to other 
geometries is discussed.  
\end{abstract}


\def\real{{\rm I\kern-.2em R}}
\def\complex{\kern.1em{\raise.47ex\hbox{
	    $\scriptscriptstyle |$}}\kern-.40em{\rm C}}
\def\integer{{\rm Z\kern-.32em Z}}
\def\pinteger{{\rm I\kern-.15em N}}
\newcommand{\arccosh}{\rm{arccosh}}
\newcommand{\arcsinh}{\rm{arcsinh}}
\newcommand{\Ai}{\rm{Ai}}
\newcommand{\Bi}{\rm{Bi}}
\newcommand{\ta}{\eta}
\newcommand{\Tr}{\rm{Tr}}
\newcommand{\im}{\rm{Im}}
\newcommand{\re}{\rm{Re}}

\narrowtext

\section{Introduction.}

The aim of this work is to present analytical methods to calculate 
the magnetic response of non-interacting electrons constrained 
to a domain with boundaries and submitted to a uniform magnetic field.

Historically, this problem traces back to the Bohr-van Leeuwen theorem 
stating the absence of classical orbital magnetism due to the exact 
cancelation between the bulk and edge magnetizations \cite{leeuwen}. Later on, 
Landau using a quantum approach did show the existence of 
a finite magnetization \cite{landau}. An extension to finite systems was 
provided by Teller \cite{teller} who showed that the Landau magnetization 
results from an almost cancelation between the bulk and edge contributions. 
In this work, we shall concentrate on the problem of non-interacting 
electrons using the semi-infinite plane as a paradigm (our methods can be
readily extended to other integrable systems - for example a disc).

Whereas it is important to know the spectrum with a sufficient precision
in order to describe low-temperature and high magnetic field phenomena 
(such as the Integer Quantum Hall Effect), 
the high temperature or weak magnetic field response like for instance   
the orbital diamagnetism is determined using 
smoothed spectral quantities. We shall describe them here by defining and 
calculating the heat kernel or equivalently its Laplace 
transform. The small time asymptotic expansion of the heat kernel is simply
related to the smooth part of the density of states (the Weyl expansion) 
\cite{Berry Howles} and to smoothed thermodynamical quantities like the 
magnetization \cite{sondheimer}. 
This asymptotic expansion of the heat 
kernel for the 
semi-infinite plane gives the perimeter correction to the Landau 
diamagnetism as noted by Robnik \cite{robnik}. We shall expand the heat 
kernel
in two different ways obtaining the same results which differ however from 
those obtained by
 Robnik. We subsequently compare our results with those obtained by using 
the Balian-Bloch \cite{balian bloch} expansion. 
We show (in agreement with more recent works \cite{john}) that the 
Balian-Bloch  method is not 
very convenient in the presence of a magnetic field due to the fact that 
several terms in the multiple scattering expansion have to be included
in order to obtain just one term in the asymptotic series of the heat kernel.
Another asymptotic method we use, following Stewartson
and Waechter \cite{stewartson}, is powerful enough to give in principle 
all the terms in the 
asymptotic series for the semi-infinite plane in a magnetic field. We argue
that all the coefficients in these series are universal in a sense that we 
shall precise.

\section{Generalities on the resolvent and the heat kernel.}

The resolvent $G(E)$ of a bounded system is by definition the Laplace transform of the heat
kernel $P(t)=\Tr\,e^{-\frac{1}{\hbar} \hat{H}t}=\sum_{n}\,e^{-\frac{1}{\hbar}E_{n}t}$, i.e.
formally:
\begin{equation}   \label{resolvent}  
G(E)=\Tr\,\frac{1}{E+\hat{H}}=\sum_{n}\,\frac{1}{E+E_{n}}, \end{equation}
where the sum runs over all eigenstates $n$ of the Hamiltonian $\hat{H}$ of the system.
This function was extensively studied for the problem of the Laplacian on 
manifolds with boundaries \cite{kac,singer}. As is well-known, the small $t$ 
behaviour of the heat
kernel of a two-dimensional system is given by Weyl's formula: 
$P(t) = \frac{mS}{2\pi\hbar t} + O(t^{-\frac{1}{2}})$, where $S$ is the 
area of the system; therefore its Laplace
transform is actually not well-defined. A natural way to regularize it is to 
subtract this leading small $t$ behaviour to $P(t)$ and then take the Laplace
transform:
\begin{equation} \label{regularized resolventbis}
g(E) = \frac{1}{\hbar} \int\limits_0^{\infty} dt (P(t)- \frac{mS}{2\pi\hbar t} ) 
e^{-\frac{1}{\hbar} Et}, \,\,\, \re (E) >0.
\end{equation}
An expression for $g(E)$ valid also for $\re (E) \leq 0$ was given 
by Berry and Howls \cite{Berry Howles}:
\begin{equation}  \label{regularized resolvent} 
g(s=\sqrt{E})=\lim_{N\to\infty}\,\left[\sum_{n\leq N} \frac{1}{s^2+E_{n}}-
\frac{mS}{2\pi\hbar^2} \ln\left\{ \frac{E_{N}}{s^{2}} \, \right\} \right]. 
\end{equation}
It is easy to show that $g(s)$ is related to the density and to the 
integrated density of states by:
\begin{eqnarray}   
\label{density of states}  
d(E) &  = & \sum_{n}\delta\left(E-E_{n}\right)=\frac{mS}{2\pi\hbar^2}-
\frac{1}{\pi}\,\lim_{\varepsilon\to0^+}\,\im\,g(i\sqrt{\,E-i\varepsilon}) \\
\label{integrated density of states}
N(E) & = & \int\limits_0^E  d(\epsilon) \, d\epsilon = \frac{mSE}{2\pi \hbar^2}
-\oint\limits_{C(E)} \frac{d\epsilon}{2i\pi}\,g(s=i\sqrt{\epsilon}),
\end{eqnarray}
where the contour $C(E)$ in the complex plane encloses the segment $[0,E]$ of the real axis
(so that it contains inside a finite number of poles of $g(i\sqrt{\epsilon})$). 
Using (\ref{density of states}) and the fact that $P(t)$ is the Laplace transform of $d(E)$,
one can show that the two definitions (\ref{regularized resolventbis})
and (\ref{regularized resolvent}) coincide for $\re (E)>0$.

Given one of the three asymptotic expansions for small
$t$ or large $s$, $E$, of $P(t)$, $g(s)$ or $d(E)$, one immediately gets the two others. 
For example, assuming the Weyl expansion of the resolvent is known:
\begin{equation}  \label{asymptotic regularized resolvent} 
g(s)\sim\sum_{r=1}^{\infty}\,\frac{c_{r}}{s^{r}}, \,\,\,s \rightarrow \infty,\,\,
| {\rm{Arg}}(s) | \leq \frac{\pi}{2}-
\Delta,\,\Delta>0,
\end{equation}
we obtain: 
\begin{eqnarray}  \label{asymptotic density of states}
d(E)  & \sim   & \frac{mS}{2\pi\hbar^2}+
\frac{1}{\pi\sqrt{\,E}}\,\sum_{r=1}^{\infty}\,\frac{(-1)^{r}}
{E^{r}}\,c_{2r+1}, \,\,\, E \rightarrow \infty \\
\label{asymptotic heat kernel}
P(t) & \sim  & \frac{mS}{2\pi \hbar t} + \sum_{r=1}^{\infty} \, \frac{c_r}
{\Gamma(\frac{r}{2})} (\frac{t}{\hbar})^{\frac{r}{2}-1}, \,\,\,t \rightarrow 0.
\end{eqnarray}
The equivalence between (\ref{asymptotic regularized resolvent}), 
(\ref{asymptotic density of states}) and (\ref{asymptotic heat kernel}) is a consequence
of equation (\ref{density of states}), the 
generalized Watson's lemma, and of its reciprocal \cite{wong}.

As a simple example, the heat kernel and the resolvent associated to the Landau spectrum of a 
particle moving on a two-dimensional
plane in a perpendicular field are easily found to be
\begin{equation} 
P_{\infty}(t) = \nonumber \frac{N_{\Phi}}{2\sinh\frac{\omega t}{2}} 
\end{equation}
and
\begin{equation}
g_{\infty}(s) = \frac{N_{\Phi}}{\hbar \omega} (\,\ln(-\nu-\frac{1}{2}) - 
\psi(-\nu)\,),
\end{equation}
if the degeneracy of the levels is $N_{\Phi}={ SB}/{ \Phi_{0}}$ and
$\nu = -\frac{s^2}{\hbar\omega}-\frac{1}{2}$ 
($\psi(z)=\frac{\Gamma '(z)}{\Gamma(z)}$ is the digamma function). 
Their asymptotic expansions are: 
\begin{eqnarray}   \label{p(t)infty asymptotic}
P_{\infty}(t) & \sim & \nonumber \frac{N_{\Phi}}{\omega t}\,(\,1-\frac{{({\omega}t})^2}
{24} + \frac{7(\omega t)^4}{5760} + \ldots ) \\
g_{\infty}(s) & \sim & \nonumber \frac{N_{\Phi}}{\hbar \omega}\,(\,-\frac{1}{24}
(\frac{\hbar \omega}{s^2})^2 + \frac{7}{960} (\frac{\hbar \omega}{s^2})^4 + \ldots ) \\
d_{\infty}(E) & \sim & \frac{N_{\Phi}}{\hbar \omega} \,\,\,\,\,\,{\rm{(no\,\,\, more\,\,\, 
terms)}}.  
\end{eqnarray}
The last result is obvious because smoothing out the regular step-like 
integrated density of states over the energies a linear function with no 
higher power-law corrections is obtained.
Because for any two-dimensional system of surface area  $S$ in magnetic field,
 $P(t)$ has same leading small 
$t$ behaviour as $P_{\infty}(t)$, another way to regularize 
the resolvent is to subtract $P_{\infty}(t)$ from $P(t)$ and then take the 
Laplace 
transform. Such a regularization gives $g(s)-g_{\infty}(s)$, and can be 
calculated in the following way. 

Let us define the Wick-rotated retarded Green's functions $G^{+}(t;{\bf{r}},{\bf{r'}})$
and $G_{\infty}^{+}  (t;{\bf{r}},{\bf{r'}})$ by $G^{+}(t;{\bf{r}},{\bf{r'}}) = 
G_{\infty}^{+}(t;{\bf{r}},{\bf{r'}}) = 0$ if $t<0$, and
\begin{equation}
(\hat{H} + \hbar \frac{\partial}{\partial t} )G^{+} (t;{\bf{r}},{\bf{r'}}) = \hbar \delta(t)
\delta({\bf{r}}-{\bf{r'}}),
\end{equation}
with the additional Dirichlet condition for $G^{+}(t;{\bf{r}},{\bf{r'}})$ on the boundary 
$\partial S$ of the system:
\begin{equation} \label{green function b c}
G^{+}(t;{\bf{r}},{\bf{r'}}) = 0 \,\,\,{\rm{for}}\,\,{\bf{r}} \in \partial S. 
\end{equation}
The Laplace transforms of these 
Green's functions, $G(E;{\bf{r}},{\bf{r'}})$ and $G_{\infty}(E;{\bf{r}},{\bf{r'}})$, 
are defined by the same boundary condition (\ref{green function b c}) and the equation:
\begin{equation}  \label{resolvent equation}
 (\hat{H} + E)G(E;\bf{r},\bf{r}\,'\,)=\delta(\bf{r}-\bf{r}\,'\,). 
\end{equation}
Then using (\ref{regularized resolventbis}) and the definition of the heat kernel, we have:
\begin{equation} \label{formula resolvent}
g(s)-g_{\infty}(s) = \int\limits_{S} d^2{\bf{r}} (\,G(s^2;{\bf{r}},{\bf{r}})
- G_{\infty}(s^2;{\bf{r}},{\bf{r}})\, ).
\end{equation}

\section{Expression of the resolvent for the semi-infinite plane.}

We propose to apply formula (\ref{formula resolvent}) to extend for the 
magnetic case the method developed by Stewartson and Waechter 
\cite{stewartson}. 
We consider the following setup. A spinless particle of charge $-e$ ($e>0$) 
and mass $m$ moves in the semi-infinite plane.
A uniform magnetic field $B$ is applied perpendicular to its surface. 
Cartesian coordinates are 
defined such that the $x$ axis is perpendicular to the boundary and the 
motion is confined to positive values of $x$, $y$ axis is along the boundary. 
Dirichlet boundary condition
is imposed at $x=0$. 
In order to work with a finite system, we introduce 
another Dirichlet boundary condition at $x=L_{\bot}$,
where $L_{\bot}$ is taken so large with respect to the magnetic length 
$l_B=\sqrt{{\hbar c}/{eB}}$ 
that an electron in an eigenstate cannot feel the presence of both boundaries. 
In this case it is clear that the contributions of the two boundaries 
to $g(s)$ will be identical (we will check this explicitly below). 
For the same reason of working with a finite system, we impose 
periodic boundary condition in the $y$ direction, taking $L$ to be the length
of the boundary.

In the Landau gauge ${\bf{A}}=(0\,,Bx)$, the 
Hamiltonian of the particle is:
\begin{equation} \label{quantum hamiltonian}
{\hat{H}}=\frac{1}{2m}\,\left( {{\hat{p}}_{x}}^{2}+({\hat{p}}_{y}+
\frac{e}{c}Bx)^{2}
\right). \end{equation}
The Green's functions $G(E;\bf{r},\bf{r}\,'\,)$ and $G_{\infty}(E;\bf{r},\bf{r}\,'\,)$
being periodic in $y-y'$ one can expand them as Fourier series, e.g.:
\begin{equation} \label{fourier transform}
G(E;{\bf{r}},{\bf{r}}\,'\,)= \sum_{p_y}
e^{\frac{ip_y (y-y\,'\,)}{\hbar}}\, G_{p_y}(E;x,x\,'\,). 
\end{equation}
The sum runs over all $p_y=n_{y}h/L$, where $n_{y} \in \integer$. 
Introducing the dimensionless variables 
$\tilde{x}={\sqrt{2} \,x}/{l_B}$, 
$\tilde{x}_0={\sqrt{2} \,p_{y}}/{m\omega l_B}$,
$N_{\Phi}'= {m\omega L^2}/{2\pi \hbar}$ and
$\nu= -\frac{E}{\hbar \omega} - \frac{1}{2}$, 
(\ref{resolvent equation}) transforms into 
\begin{equation} \label{transformed equation} 
\left( \frac{d^{2}}{d\tilde{x}^{2}} - \frac{E}{\hbar\omega} - 
\frac{1}{4}\,(\tilde{x}+\tilde{x}_0)^{2} \right) \,  G_{p_y}(E;\tilde{x},
\tilde{x}') = 
-\frac{2m}{\hbar^2} (4\pi N_{\Phi}')^{-\frac{1}{2}} \delta 
(\tilde{x}-\tilde{x}'). 
\end{equation}
The solution of this equation such that 
$G_{p_y}(E;0,x'\,)= G_{p_y}(E;L_{\bot},x'\,)= 0$ can be 
written as a sum of three terms:
\begin{equation} \label{rrr}
G_{p_y}(E;x,x'\,) =  
-\frac{2m}{\hbar^2 \sqrt{4\pi N_{\Phi}'}} ( \tilde{G}_{\infty} (\nu,\tilde{x}_0;\tilde{x},\tilde{x}')
+ \tilde{G}_{1b}(\nu,\tilde{x}_0;\tilde{x},\tilde{x}')  
+ \tilde{G}_{2b}(\nu,\tilde{x}_0;\tilde{x},\tilde{x}')  ).
\end{equation}
The Green's function for unbounded motion 
$\tilde{G}_{\infty} (\nu,\tilde{x}_0;\tilde{x},\tilde{x}')$ is given by:
\begin{equation}
\tilde{G}_{\infty} (\nu,\tilde{x}_0;\tilde{x},\tilde{x}')  = 
-\frac{\Gamma(-\nu)}{\sqrt{2\pi}}
D_{\nu} (\tilde{x}_{>}- \tilde{x}_0) D_{\nu}(\tilde{x}_0- \tilde{x}_{<}), 
\end{equation}
where $\tilde{x}_{>}= {\rm{max}} \{ \tilde{x},\tilde{x}' \}$, 
$\tilde{x}_{<}= {\rm{min}} \{ \tilde{x},\tilde{x}' \}$ 
(we used the Wronskian \cite{buchholz} 
$W(\,D_{\nu}(u), D_{\nu}(-u)\,) = \frac{\sqrt{2\pi}}{\Gamma(-\nu)}$).
The two last terms in (\ref{rrr}) are respectively the contributions of the 
boundaries $x=0$ and $x=L_{\bot}$ to the Green's function. They are found by 
requiring that $G_{p_y}(E;x,x'\,)$ have to vanish on the
boundaries and neglecting exponentially small terms due to the asymptotic
behaviour of the function $D_{\nu}$ as its argument is large and positive
(recall that $\tilde{L}_{\bot} ={\sqrt{2}L_{\bot} }/{l_B} \gg 1$):
\begin{eqnarray}
\tilde{G}_{1b}(\nu,\tilde{x}_0;\tilde{x},\tilde{x}) &=& \nonumber
\frac{\Gamma(-\nu)}{\sqrt{2\pi}} \frac{D_{\nu}(\tilde{x}_0)}{D_{\nu}(-\tilde{x}_0)}
{D^2_{\nu}(\tilde{x}-\tilde{x}_0)} \\
\tilde{G}_{2b}(\nu,\tilde{x}_0;\tilde{x},\tilde{x}) &=& 
\frac{\Gamma(-\nu)}{\sqrt{2\pi}} \frac{D_{\nu}(\tilde{L}_{\bot} -\tilde{x}_0)}
{D_{\nu}(\tilde{x}_0-\tilde{L}_{\bot} )} D^2_{\nu}(\tilde{x}_0-\tilde{x}). 
\end{eqnarray}
Since we are interested in the limit $L \rightarrow \infty$, we can now 
replace the sum over 
$p_y$ in (\ref{fourier transform}) by an integral. We deduce that 
$\tilde{G}_{1b}(\nu,\tilde{x}_0;\tilde{x},\tilde{x})$ and 
$\tilde{G}_{2b}(\nu,\tilde{x}_0;\tilde{x},\tilde{x})$ give 
the same contribution to 
(\ref{formula resolvent}), which can be expressed
using the relevant formulas \cite{buchholz} for the integrals of the 
function $D_{\nu}$ as:
\begin{eqnarray}
\label{resolvent2}  \nonumber
g(s)-g_{\infty}(s) & = & \frac{\Gamma(-\nu)\sqrt{N_{\Phi}'} }{\hbar
\omega\sqrt{8\pi^2}}
\intop_{-\infty}^{\infty}d\tilde{x}_{0}\frac{D_{\nu}({\tilde{x}}_{0})}
{D_{\nu}(-{\tilde{x}}_{0})} \left(D_{\nu}(-{\tilde{x}}_{0})
\frac{\partial D_{\nu}'}{\partial\nu}(-{\tilde{x}}_{0})  -
D_{\nu}'(-{\tilde{x}}_{0}) \frac{\partial D_{\nu}}{\partial
\nu}(-{\tilde{x}}_{0})
\right)  \\
& = & \frac{\sqrt{N_{\Phi}'} }
{\sqrt{4\pi} \hbar \omega} \lim_{\tilde{X}_0 \rightarrow \infty} \left(
 \,\frac{\partial}{\partial\nu}
\intop_{-\tilde{X}_0}^{\tilde{X}_0} d\tilde{x}_0 \ln D_{\nu}(-\tilde{x}_0)
-\tilde{X}_0 \psi(-\nu) \,
\right).
\end{eqnarray}
This is the resolvent for the semi-infinite plane. In the infinitely large 
strip geometry, $g(s)-g_{\infty}(s)$ is twice this result.

Formula (\ref{resolvent2}) can be understood more intuitively as follows. 
The integrated density of states $N(E)$
count the number of zeros of the wavefunction at the origin $D_{\frac{\epsilon}{\hbar\omega}-
\frac{1}{2} }(\tilde{x}_0)$, for all $\tilde{x}_0$ and all energies $\epsilon$ up to $\epsilon=E$:
\begin{equation} \label{integrated dos2}
N(E)=\sum_{p_y}\,\oint\limits_{C(E)} \frac{d \epsilon}{2\pi{i}} \frac{\partial}{\partial\epsilon}
\ln D_{\frac{\epsilon}{\hbar\omega}-\frac{1}{2}}(\tilde{x}_0).
\end{equation}
The sum and the integral cannot be inverted since there are infinitely many states. However,
we can formally subtract from $N(E)$ the integrated density of states of the infinite plane 
which we write as:
\begin{equation}
N_{\infty}(E)= \frac{1}{2} \sum_{p_y} \sum_{n_x} \theta(\,E- \hbar \omega (n_x+\frac{1}{2})\,) 
= -\frac{1}{2\hbar\omega} \sum_{p_y} \,\oint\limits_{C(E)} \frac{d \epsilon}{2\pi{i}} 
\psi (-\frac{\epsilon}{\hbar\omega}+\frac{1}{2}).
\end{equation}
(a factor $\frac{1}{2}$ is introduced since we neglected the second boundary 
contribution in (\ref{integrated dos2})).
Using equation (\ref{integrated density of states}) we obtain:
\begin{equation} \label{resolvent3}
g(s)- g_{\infty}(s) =  \nonumber \frac{1}{\hbar\omega} \sum_{p_{y}} (\, \frac{\partial}
{\partial\nu} \ln D_{\nu}(\tilde{x}_0) -\frac{1}{2} \psi (-\nu)\,) % \\
%& = & \frac{\sqrt{N_{\Phi}'} }
%{\sqrt{4\pi} \hbar \omega} \lim_{\tilde{X}_0 \rightarrow \infty} \{ \,\frac{\partial}{\partial\nu}
%\intop_{-\tilde{X}_0}^{\tilde{X}_0} d\tilde{x}_0 \ln D_{\nu}(-\tilde{x}_0) -
%\tilde{X}_0 \psi
%(-\nu) \,\}. 
\end{equation}
Replacing here the sum over $p_{y}$ by an integral we obtain Eq. (\ref{resolvent2}).
The heat kernel is found by performing inverse Laplace transform 
on (\ref{resolvent2}) or (\ref{resolvent3}).

It is interesting to note that $g(\sqrt{E})$ and $P(t)$ have the following 
forms:
\begin{eqnarray} \label{scale argument} \nonumber
\hbar \omega g(\sqrt{E}) & = & N_{\Phi}\,\tilde{g}_{\infty}(\frac{E}
{\hbar\omega}) + 
\sqrt{N_{\Phi}'} \,\tilde{g}_S(\frac{E}{\hbar\omega})  \\ 
P(t) & = &  N_{\Phi} \,\tilde{P}_{\infty}(\omega t)+ \sqrt{N_{\Phi}'} 
\,\tilde{P}_S(\omega t),
\end{eqnarray}
where $N_{\Phi}= {m \omega L L_{\bot} }/{2 \pi \hbar}$, 
$N_{\Phi}'= {m \omega L^2 }/
{2 \pi \hbar}$ and none of the functions decorated by tilde depend explicitly 
on  $B$, $L$ or $L_{\bot}$.
The boundary term is proportional to $\sqrt{N_{\Phi}'}$ and smaller than
the bulk term by a factor of order $\frac{1}{\sqrt{N_{\Phi}} }$ for 
$L_{\bot} \sim L$. For finite lengths $L$ of the boundary, there are 
exponentially small correction terms 
in (\ref{resolvent2}) and 
(\ref{scale argument}) due to the error introduced by replacing 
the sum in 
by an integral in (\ref{fourier transform}). Neglecting these small 
corrections, an asymptotic 
expansion of the heat kernel and of the density of states in  
$h_{eff} = {1}/{N_{\Phi}'}$, which is a small parameter 
(assuming large fields or large systems), has two terms (this is
due to the straight boundary - there will be more terms if an edge has 
a non-zero curvature).

The arguments of this section extend to any separable system provided one can 
reduce the problem to
a one-dimensional Sturm-Liouville problem. As an example we 
derive in the appendix the resolvents of a particle moving in a disc
 with and without a magnetic field.

\section{Weyl expansions of the resolvent and the heat kernel.}  

In this section we shall derive asymptotic expressions for the heat kernel 
and the resolvent
using equation (\ref{resolvent3}). For large positive energies 
(large negative $\nu$) and large $\tilde{x}_{0}$ the Darwin asymptotic 
expansion (basically WKB expansion of the wavefunction)
of the parabolic cylinder function 
\cite{darwin,abramowitz} is
\begin{equation}  \label{darwin expansion}
{\ln}D_{\nu}({\tilde{x}}_{0})\sim \frac{{\ln}2\pi}{4}-\frac{1}{2}\ln
\Gamma(-\nu) -\theta({\tilde{x}}_{0},a) -\frac{1}{4}{\ln}({\tilde{x}}_{0}
^{2}+4a) +\sum_{s=1}^{\infty}\frac{(-1)^{s}d_{3s}}{(\sqrt{{\tilde{x}}_{0}
^{2}+4a})^{3s}},
\end{equation}
where $\theta({\tilde{x}}_{0},a)$ is an odd function of $\tilde{x}_0$, 
$a=\frac{E}{\hbar\omega}>0$ and ${\tilde{x}}_{0}^{2}+4a \gg 1$.
The coefficients $d_{3s}$ are odd functions of ${\tilde{x}}_{0}$ for odd values of
$s$ and even functions of ${\tilde{x}}_{0}$ for even $s$. Since we integrate ${\tilde{x}}_{0}$ in
(\ref{resolvent3}) over a symmetric interval, only even functions of 
${\tilde{x}}_{0}$ contribute. The first three even-indexed 
coefficients $d_{3s}$ are given respectively by \cite{darwin}:
$d_{6}  =  \frac{3}{4}{\tilde{x}}_{0}^{2}-2a$, 
$d_{12} = \frac{153}{8}{\tilde{x}}_{0}^{4}-186a{\tilde{x}}_{0}^{2}+80a^2$ 
and $d_{18} =  \frac{6381}{4}{\tilde{x}}_{0}^{6}-29862a
{\tilde{x}}_{0}^{4} +62292a^2 {\tilde{x}}_{0}^{2}-\frac{31232}{3}a^3$. \\
Using (\ref{resolvent3}) and (\ref{p(t)infty asymptotic}) the asymptotic 
expansion of the resolvent is obtained:
\begin{eqnarray}  \label{expansion2}
g(s=\sqrt{\hbar\omega a}) & \sim & \nonumber \frac{N_{\Phi}}{\hbar\omega} (\,-\frac{1}{24a^2}
+\frac{7}{960a^4} + \ldots ) \\
& & - \frac{\sqrt{\pi N_{\Phi}'}}{4\hbar\omega} (\frac{1}{\sqrt{a}}- 
\frac{9}{256 a^{\frac{5}{2}}}+\frac{2625}{262144 a^{\frac{9}{2}}}-
\frac{241197}{2^{25} a^{\frac{13}{2}}}+{\ldots} ).
\end{eqnarray}
The Weyl expansion of the heat kernel is derived using 
(\ref{asymptotic heat kernel}) 
\begin{equation}  \label{weyl2}
P(t= \frac{\tau}{\omega})= \frac{N_{\Phi}}{\tau}(1-\frac{\tau^2}{24}+\frac{7\tau^4}{5760}-\ldots)-
\frac{\sqrt{N_{\Phi}'}}{4\sqrt{\tau}} (1-\frac{3\tau^
{2}}{64}+\frac{25\tau^{4}}{16384}-\frac{7309\tau^6}{315\,\, 2^{20}} + {\ldots}).
\end{equation}
These expansions could in principle be continued indefinitely, calculating 
recursively the coefficients ${d_{6s}}$.

It is instructive to compare the expression (\ref{weyl2}) with its
counterpart at $B=0$. The heat kernel in the absence of magnetic field is
simply obtained from (\ref{weyl2}) by taking the argument of the analytic 
function of $\tau$ in each parenthesis to be zero. 
This suggests that a natural
extension of the heat kernel expansion for the two-dimensional shape without
the magnetic field 
given for example in \cite{Berry Howles,stewartson} to the case of magnetic 
billiard is to multiply each term of that expansion 
by an analytic function of $\tau$. These
functions should be universal for all flat two-dimensional billiards with
smooth boundaries.

\section{The Balian-Bloch method.} 

The small time asymptotic expansion of the heat kernel can also be found 
using a method suggested by  Balian and Bloch \cite{balian bloch}.
It consists in a reformulation of the 
problem of solving a partial differential equation of elliptic type 
with Dirichlet (or another) boundary condition in terms of an integral 
equation of the Fredholm type.
 This integral equation is then solved iteratively 
(Neumann series), and each term in this multiple reflection expansion 
corresponds to one term in the asymptotic series of the heat
kernel, as shown by Balian and Bloch \cite{balian bloch}. 

The Green's function for the Dirichlet problem in a 
domain $S$ with a boundary $\partial S$ in absence of magnetic field is written as a sum 
of two terms:
\begin{equation}  \label{bb green function}
G(E;{\bf r},{\bf r}'\,) = G_{\infty}(E;{\bf r},{\bf r}'\,) + 
G_{S}(E;{\bf r},{\bf r}'\,), 
\end{equation}
where each term in the right-hand side satisfy (\ref{resolvent equation}) in which $\hat{H}=
-\frac{\hbar^2}{2m} \Delta$ is the Laplacian. The first term is the infinite 
plane Green's function; the second is specified by  
the boundary condition: $G_{S}(E;{\bf r},{\bf r}'\,)=- G_{\infty}(E;{\bf r},{\bf r}'\,)$ 
for  ${\bf r}$ on $\partial S$.
This boundary term is expressed in terms of an unknown density $\mu_{E}
(\mbox{\boldmath$\alpha$},{\bf r})$ as:
\begin{equation} \label{bb boundary term}
G_{S}(E;{\bf r},{\bf r}'\,)=\intop_{\partial S}d{\sigma_{\alpha}}
\frac{\partial G_{\infty}(E;{\bf r},\mbox{\boldmath$\alpha$})}{\partial 
n_{\alpha}}  \mu_{E}(\mbox{\boldmath$\alpha$},{\bf r}'\,), \end{equation}
and $\mu_{E}(\mbox{\boldmath$\alpha$},{\bf r})$ is determined by 
solving the following Fredholm integral equation:
\begin{equation} \label{bb fredholm} 
\frac{m}{\hbar^2} \mu_{E}(\mbox{\boldmath$\alpha$},{\bf r}'\,) = - 
G_{\infty}(E;\mbox{\boldmath$\alpha$},{\bf r}'\,)-\intop_{\partial S}d{\sigma_{\beta}}
\frac{\partial G_{\infty}(E;\mbox{\boldmath$\alpha$},\mbox{\boldmath$\beta$})}
{\partial n_{\beta}} \mu_{E}(\mbox{\boldmath$\beta$},{\bf r}'\,), 
\end{equation}
where $\mbox{\boldmath$\alpha$}, \mbox{\boldmath$\beta$}, \ldots $ are 
arbitrary points on the boundary $\partial S$, $d{\sigma_{\beta}}$ is the boundary 
differential element, and $\frac{\partial}{\partial n_{\beta}}$ is the normal 
derivative at the point $\mbox{\boldmath$\beta$}$ with the normal oriented 
towards the interior of the domain. Solving iteratively the integral 
equation (\ref{bb fredholm}) for the density $\mu_{E}(\mbox{\boldmath$\alpha$},
{\bf r})$, the following 
multiple reflection expansion is obtained for
the Green's function:
\begin{eqnarray}    \label{bb neumann}                                         
G(E;{\bf r},{\bf r}'\,) = G_{\infty}(E;{\bf r},{\bf r}'\,) - \frac{\hbar^2}{m}
\intop_{\partial S}d{\sigma_{\alpha}} \frac{\partial G_{\infty}(E;{\bf r},
\mbox{\boldmath$\alpha$})} {\partial n_{\alpha}}  G_{\infty} \nonumber
(E;\mbox{\boldmath$\alpha$},{\bf r}'\,) + \\ \left( \frac{\hbar^2}{m} 
\right)^{2} \intop_{\partial S}d{\sigma_{\alpha}}d{\sigma_{\beta}}
\frac{\partial G_{\infty}(E;{\bf r},\mbox{\boldmath$\alpha$})}{\partial 
n_{\alpha}}  \frac{\partial G_{\infty}(E;\mbox{\boldmath$\alpha$},
\mbox{\boldmath$\beta$})}{\partial n_{\beta}}  G_{\infty}
(E;\mbox{\boldmath$\beta$},{\bf r}'\,) -\ldots  \end{eqnarray}
By mean of the time-dependent Green's function 
$G^{+}(t;{\bf r},{\bf r}')$ introduced in section 4, we can calculate the heat kernel: 
$P(t)= \int_{S} d^2{\bf{r}} G^{+}(t;{\bf r},{\bf r})$.
The multiple expansion of $G^{+}(t;{\bf r},{\bf r}')$ is: 
\begin{eqnarray}     \label{time neumann}    
G^{+}(t;{\bf r},{\bf r}'\,) = G^{+}_{\infty}(t;{\bf r},{\bf r}'\,) - 
\frac{\hbar}{m}
\intop_{\partial S}d{\sigma_{\alpha}}\intop_{0}^{t}d\tau \nonumber
\frac{\partial G^{+}_{\infty}(t-\tau;{\bf r},\mbox{\boldmath$\alpha$})}
{\partial n_{\alpha}}  G^{+}_{\infty}(\tau;\mbox{\boldmath$\alpha$},{\bf r}'\,) + 
\\  \left( \frac{\hbar}{m} \right)^{2} \intop_{\partial S}d{\sigma_{\alpha}}
d{\sigma_{\beta}}  \intop_{0}^{t}d\tau_1\intop_{0}^{\tau_1}d\tau_2
\frac{\partial G^{+}_{\infty}(t-\tau_1;{\bf r},\mbox{\boldmath$\alpha$})}
{\partial n_{\alpha}} \frac{\partial G^{+}_{\infty}(\tau_1-\tau_2;
\mbox{\boldmath$\alpha$},\mbox{\boldmath$\beta$})} {\partial n_{\beta}}  
G^{+}_{\infty}(\tau_2;\mbox{\boldmath$\beta$},{\bf r}'\,) -\ldots
\end{eqnarray}

This approach was applied in this form to the case with a
uniform magnetic field perpendicular to the domain $S$ \cite{robnik}. However, 
 at each order of the obtained multiple reflection expansion we obtain a 
term which is not gauge invariant.
 This problem may be easily corrected by 
introducing the covariant derivative $\frac{\partial}{\partial n_{\alpha}}-
\frac{ie}{{\hbar}c}A_{n}(\mbox{\boldmath$\alpha$})$ instead of the usual 
$\frac{\partial}{\partial n_{\alpha}}$. This substitution is of no importance
when the gauge is chosen such that the vector potential has no component
 normal to the boundary (as it happens to be in our problem), but generally
should be taken into account.

Let us therefore apply the Balian and Bloch method to the
 semi-infinite plane in a uniform magnetic field.
The Green's function for the infinite plane is given by:
\begin{equation}  \label{time infty green}
G^{+}_{\infty}(t;{\bf r},{\bf r}'\,) = \frac{m\omega}{4\pi\hbar\sinh
\frac{{\omega}t}{2}} e^{-\frac{m\omega}{4\hbar}({\bf r}-{\bf r}'\,)^2\coth
\frac{{\omega}t}{2}-\frac{im\omega}{2\hbar}(y-y'\,)(x+x'\,)}. \end{equation} 

We calculate now the first term (proportional to $B^2$) 
in the small magnetic field expansion of the 
heat kernel. It turns out that the first three 
boundary-dependent terms in the multiple reflection expansion do contribute to 
this order. The small $B$ expansion of the one-reflection term begins with:
\[-\frac{L}{8}\sqrt{\frac{2m}{\pi{\hbar}t}}+
  \frac{L}{8}\sqrt{\frac{2m}{\pi{\hbar}t}}\frac{7}{192}{\omega^2}t^2,  \] 
for the two-reflection and three-reflection terms, it is respectively:
\[ \frac{L}{8}\sqrt{\frac{2m}{\pi{\hbar}t}}\frac{3}{192}{\omega^2}t^2,  \] 
and
\[  -\frac{L}{8}\sqrt{\frac{2m}{\pi{\hbar}t}}\frac{1}{192}{\omega^2}t^2.  \] 
Therefore the small magnetic field expansion of the heat kernel begins as 
follows:
\begin{equation} \label{bb heat kernel expansion} 
P(t)=P_{\infty}(t)-\frac{L}{8}\sqrt{\frac{2m}{\pi{\hbar}t}}+
  \frac{L}{8}\sqrt{\frac{2m}{\pi{\hbar}t}}\frac{3}{64}{\omega^2}t^2-\ldots 
\end{equation} 
in accordance with the previously obtained result (\ref{weyl2}). This however 
disagrees with Robnik's calculation \cite{robnik}. In this work the author 
assumed that in the case
of a zero-curvature boundary, the one-reflection term contains all the 
correction due to the boundary (as it is for the problems with zero magnetic
field). However as already  noted by John and Suttorp \cite{john}, higher 
order terms can not be neglected in calculating physical quantities in the 
geometry with straight boundaries in the presence of a magnetic field. We 
also remark that in order
to calculate the term proportional to ${\omega}^{2n}$ in the asymptotic 
expansion of the heat kernel, all the multiple reflection terms up to the 
$2n+1$'th order have to be taken into account. This result is intuitively
appealing. Indeed as the magnetic field increases, the trajectories of the
particles bend more and more, so that higher and higher terms in the 
multiple reflection expansion do contribute.

\section{Perimeter corrections to the Landau diamagnetism.}

We apply in this section the results of the preceding sections to compute the
magnetic susceptibility of an independent electron gas on a strip of very 
 large width $L_{\bot}$.
 
Let us first consider the case of a non-degenerate gas at temperature $T=\frac{1}{k_B \beta}$. 
The heat kernel $P(t)$ for $t=\tau/\omega=
\hbar \beta$  is nothing but the one-electron partition function. 
If $N$ is the electronic density (per unit area) and $\mu_B=\frac{e\hbar}{2mc}$ the
Bohr magneton, the magnetic susceptibility of the ideal gas is given by 
\begin{equation}
\chi = \lim\limits_{\tau \rightarrow 0} 
4 N \mu_B^2 \beta \frac{\partial^2}{\partial\tau^2}{\rm{ln}} Z.
\end{equation}
Using the expansion (\ref{weyl2}) of $P(t)$ we obtain the weak field 
susceptibility, 
\begin{equation}  \label{nondegenerate}
\chi =  - \frac{1}{3} N \beta \mu_B^2 \left( 1 - \frac{\lambda_T}{16 L_{\bot}}
 (1-\frac{\lambda_T}{2L_{\bot} })^{-1}    \right),  
\end{equation}
where $\lambda_T=\sqrt{\frac{\pi\beta \hbar^2}{2m}}$ is the de Broglie 
thermal length. Like in \cite{robnik} the correction to the Landau 
diamagnetic susceptibility $\chi_{\infty}=-\frac{1}{3}N \beta \mu_B^2$ is 
paramagnetic, but is smaller by one order of magnitude. 
Our result agrees with numerical calculations \cite{ruitenbeck}.

For a degenerate electron gas of Fermi energy $E_F$, the connection between 
the susceptibility
and the partition function is different \cite{sondheimer}. In the 
grand-canonical 
ensemble, $\chi$ is given by 
\begin{equation} 
\chi = - \lim\limits_{B \rightarrow 0} \frac{1}{S} \frac{\partial^2 
\Omega(E_F)}{\partial B^2},
\end{equation}
$\,\Omega(E_F)$ being the thermodynamic potential at the 
chemical potential $E_F$. The thermodynamic potential at zero temperature 
$\,\Omega_0 (E_F)$ is connected to the partition function (heat kernel)
 through \cite{sondheimer}
\begin{equation} 
\Omega_0 (E_F)= -\int\limits_{-i\infty\,+0+}^{i\infty\,+0+} \frac{d\beta}{2\pi i}\,
e^{\beta E_F} \beta^{-2} Z(\beta).
\end{equation}
Using (\ref{scale argument}) we obtain:
\begin{equation}
\Omega_0 (E_F) = -\frac{mS}{2\pi} \int\limits_{-i\infty\,+0+}^{i\infty\,+0+} \frac{d\tau}{2\pi i}\,
e^{\frac{\tau E_F}{\hbar \omega}}
\left(  \omega^2 \tau^{-2} \tilde{P}_{\infty}(\tau) + 
   \sqrt{\frac{2\pi\hbar}
{mL_{\bot}^2} } 
\omega^{\frac{3}{2}} \tau^{-2} \tilde{P}_S(\tau)
\right).
\end{equation}
Replacing in this equation the Weyl expansions (\ref{weyl2}) of 
$\tilde{Z}_{\infty}(\tau)$ and 
$\tilde{Z}_S(\tau)$ amounts to neglecting all oscillating contributions due to 
periodic orbits in
$\Omega_0(E_F)$. However, even if the oscillating terms in 
the thermodynamic
potential can at very low temperature give the largest corrections to the 
Landau diamagnetism, these
terms are exponentially
damped by temperature \cite{jalabert,gurevich} (more precisely, the 
contribution of each periodic
orbit is damped by the factor 
$\frac{2\pi r m L_{\bot}}{\hbar^2 \beta k_F} \sinh^{-1}
(\frac{2\pi r m L_{\bot}}{\hbar^2 \beta k_F})$, where $r$ is the number of 
repetitions of the orbit
and  $k_F=\frac{\sqrt{2mE_F}}{\hbar}$ is the Fermi wavevector).
For temperatures such that $\frac{\hbar^2 k_F}{m L_{\bot}} \ll k_B T \ll E_F$, one can neglect these
oscillating contributions, and we have at small field: 
\begin{equation}  \label{degenerate}
\chi \simeq {\bar{\chi}} = - \frac{e^2}{24\pi mc^2} \left(1- \frac{9}{16} \frac{1}{L_{\bot} k_F}
\right). 
\end{equation}
Here again the correction is paramagnetic and coincides with perturbative calculations 
\cite{gurevich}. 

The expressions (\ref{nondegenerate}) and (\ref{degenerate}) 
give the magnetic susceptibilities of a semi-infinite plane regularized
to have a finite  
 width $L_{\bot}$ for small fields $B$ such that 
$L_{\bot} \gg l_B$. Here again the second 
term in these equations have to be multiplied by two if 
one wishes to consider a finite very large strip (with two boundaries) 
instead of the semi-infinite plane.

The susceptibilities (\ref{nondegenerate}) and
(\ref{degenerate}) give the perimeter corrections to the Landau diamagnetism 
for a general
macroscopic billiard of surface $S$ and smooth boundary of length $L$, 
with $L_{\bot}={S}/{L}$. 
However for a generic billiard this correction is not the only one, 
and other terms due to 
the curvature of the boundary must be added.

\section{Conclusion}

We obtained an exact relation for the resolvent and the heat kernel in the 
semi-infinite plane geometry in the uniform magnetic field. 
This relation is a convenient starting point
for the asymptotic expansion, using the WKB approximation for the 
wavefunctions. The validity of the method 
extends beyond the special case of the semi-infinite geometry,
in fact it works for any separable system. 
The main result of the work is the boundary term of the heat kernel
asymptotic expansion, which is an infinite series calculable recursively 
in principle. The first terms in this series were verified using
the Balian-Bloch multiple scattering expansion, which is less convenient
practically than the method of Stewartson and Waechter. 
We point out how the presence of the magnetic field affects the physics of
the Balian and Bloch approach making necessary to consider three terms
in the multiple scattering series in order to get correctly the weak field
response. The corrected perimeter corrections to the Landau diamagnetism
are derived.

The properties of the heat kernel of manifolds have been extensively 
investigated starting (among others) from the work of Kac
\cite{kac} in order to relate the spectral and geometrical 
descriptions \cite{singer}.
Our results could be seen as an extension of these works to the case of 
magnetic
billiards. For the case of a straight boundary, the magnetic field adds an 
infinite series to the bare perimeter term, which is, in some sense, 
equivalent to an effective curvature of the boundary. If the boundary has
a curvature, another length scale enters the problem, coupling with the
cyclotron radius.

{\bf Acknowledgement}

This work was supported in part by a grant from the Israel Science
foundation and by the fund for promotion of the research at the Technion.
R.~N. acknowledges support by BSF grant 01-4-32842.

\renewcommand{\thesection}{\Alph{section}}
\setcounter{section}{0}
%\appendix
\section{Appendix}
\renewcommand{\theequation}{\thesection.\arabic{equation}}
\setcounter{equation}{0}
Consider a particle confined to a disc of radius $R$. Introducing
the momentum $k^2=\frac{2mE}{\hbar^2}$, 
the energies are solutions of the equation
$J_{l}(kR)=0$,   
where $l$ is the angular momentum quantum number.
The number of zeros is equal to the integral of the logarithmic derivative of
the Bessel function:
\begin{equation}
N(E)= \sum\limits_l \,\oint\limits_{C(E)} \frac{d\epsilon}{2\pi{i}}\frac{\frac{d}{d\epsilon}
J_{l}(kR)}{J_{l}(kR)}.  \end{equation}
Then according to our method, we obtain the resolvent
\begin{equation}
G(E) = \sum_{l}\,\frac{m}{\hbar^2 {k}}\cdot \frac{\frac{d}{dk}I_{l}(kR)}
{I_{l}(kR)}.  \end{equation}
which gives the result of Stewartson and Waechter 
\cite{stewartson} after subtracting the resolvent corresponding to the
 infinite plane. 

In the presence of a magnetic field, the energy of the particle moving in 
a circular billiard is given by:
$_{1}F_{1}(\frac{l+|l|+1}{2} - \frac{E}{\hbar\omega};1+|l|;N_{\Phi})=0$,
where $N_{\Phi}=\frac{SB}{\Phi_0}$. The counting function is now: 
\begin{equation} 
N(E)=\sum_{l}\,\oint\limits_{C(E)} \frac{d\epsilon}{2\pi{i}}\frac{\frac{d}{d\epsilon}   
{_{1}F_{1}}(\frac{l+|l|+1}{2} - \frac{\epsilon}{\hbar\omega};1+|l|;N_{\Phi})}
{_{1}F_{1}(\frac{l+|l|+1}{2} - \frac{\epsilon}{\hbar\omega};1+|l|;N_{\Phi})},
\end{equation}
and the resolvent is: 
\begin{equation} 
G(E) = \sum_{l}\,\frac{\frac{d}{d\,E}
{_{1}F_{1}}(\frac{l+|l|+1}{2} + \frac{E}{\hbar\omega};1+|l|;N_{\Phi})}
{_{1}F_{1}(\frac{l+|l|+1}{2} + \frac{E}{\hbar\omega};1+|l|;N_{\Phi})}.
\end{equation}
We emphasize again that in order
 to work with well-defined quantities we must subtract from this 
expression the part of the resolvent corresponding to the infinite plane.

\end{document}